\documentstyle[sprocl,graphicx]{article}

\def\be{\begin{equation}}
\def\ee{\end{equation}}
\def\bed{\begin{equation}}
\def\eed{\end{equation}}
\def\bea{\begin{eqnarray}}
\def\eea{\end{eqnarray}}

\newcommand{\dn}[2]{{d}^{#1} #2 \:}
\newcommand{\Corrftn}[2]{C_{\vec{#1}} (\vec{#2})}
\newcommand{\Rftn}[2]{{\cal R}_{\vec{#1}} (\vec{#2})}
\newcommand{\Clm}[3]{C_{#1 #2} (#3)}
\newcommand{\Source}[2]{S_{\vec{#1}} (\vec{#2})}
\newcommand{\Sourcenovec}[2]{S_{#1} (#2)}

\newcommand{\wftn}[2]{\Phi^{(-)}_{\vec{#1}} (\vec{#2})}
\newcommand{\wfsquare}[2]{\left| \wftn{#1}{#2} \right|^2}
\newcommand{\spectra}[1]{\frac{dN_{1}}{d\vec{#1}}}
\newcommand{\prc}[1]{Phys.~Rev.~C {#1}}

\newcommand{\prl}[1]{Phys.~Rev.~Lett.~{#1}}
\newcommand{\plb}[1]{Phys.~Lett.~B {#1}}

\newcommand{\pdd}{P.~Danielewicz}

\def\gapproxeq{\raisebox{-.5ex}{$\,\stackrel{>}{\scriptstyle
\sim}\,$}}
\def\lapproxeq{\raisebox{-.5ex}{$\,\stackrel{<}{\scriptstyle
\sim}\,$}}

\def\etal{{\em et al.}}

\begin{document}

\title{IMAGING OF SOURCES IN HEAVY-ION REACTIONS}

\author{ P.~DANIELEWICZ\protect\footnote{Talk given at the
Int.\ Workshop
on ``Collective Excitations in Fermi and Bose Systems", Serra
Negra, Brazil, September 14--17, 1998.},
D.~A.~BROWN}

\address{
      National Superconducting Cyclotron Laboratory\\
       and Department of Physics and Astronomy\\
       Michigan State University, East Lansing, MI 48824-1321, USA}


\maketitle\abstracts{
I discuss imaging sources from low relative-velocity correlations in
heavy-ion reactions.  When the correlation is dominated by interference,
we can obtain the images by Fourier transforming the correlation
function.  In the~general case, we may use the method of optimized
discretization.  This method stabilizes the inversion by adapting the
resolution of the source to the experimental error and to the measured
velocities.  The~imaged sources contain information on freeze-out density,
phase-space density, and resonance decays, among other things.
}

\section{Introduction}

Imaging techniques are used in many diverse
areas such as geophysics, astronomy, medical diagnostics, and police
work.  The goal of imaging varies widely from determining the density
of the Earth's interior to reading license plates from
blurred photographs to issue speeding fines.  My own interest
in the problem stems from seeing an image of Betelguese, a~red
giant $\sim 600$~ly away that has irregular features changing
with time.  The~image was obtained using intensity
interferometry such as used in nuclear physics~\cite{boa90}.
After seeing this, the natural question was whether images could be
obtained for nuclear reactions.  Needless to say, answers to such
questions tend to be negative.

In a~typical imaging problem, the~measurements yield
a~function (in our case, the correlation function $C$) which is related
in a~linear fashion to the function of interest (in our case, the
source function $S$):
\bed
\label{CKS}
 C(q) = \int dr \, K(q,r) \, S(r) \, .
\eed
In other words, given the data for $C$ with errors, the task of
imaging is the determination of the source function~$S$.
Generally, this requires an~inversion of the kernel~$K$. The~more
singular the kernel~$K$, the~better the chances for a~successful
restoration of~$S$.

\begin{figure}
\begin{center}
\includegraphics[angle=0.0,
width=0.72\textwidth]{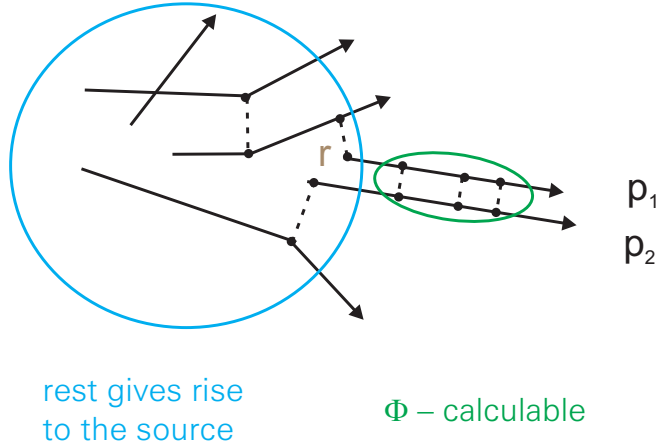}
\end{center}
\caption{Separation of the interference and final-state
interactions, in terms of the two-particle wavefunction, from
the amplitude for the reaction.}
\label{source}
\end{figure}

In reactions with many particles in the final state, there is a~linear
relation of the type (\ref{CKS})
between the two-particle cross section $d^6 \sigma / d^3 {
p}_1 \, d^3 {
p}_2$ and the unnormalized relative distribution of emission points~$S'$ for
two particles.  Interference and interaction terms between the two
particles of interest may be separated out from the general amplitude for the
reaction and described in terms of the two-particle
wavefunction~$\Phi^{(-)}$ (see Fig.~\ref{source}).
The~rest of the amplitude squared, integrated in the cross
section over unobserved particles, yields the unnormalized Wigner
function~$S'$ for the distribution of emission points written
here in the two-particle frame:
\bed
{d^6 \sigma \over d^3{ p}_{1} \, d^3{ p}_{2}} =
\int
d^3{ r} \,
S'_{\vec{P}}(\vec{ r}) \,
|\Phi_{\vec{ p}_1 - \vec{ p}_2}^{(-)} (\vec{ r})|^2 \, .
\label{2PS}
\eed
The vector $\vec{ r}$ is the relative separation between
emission
points and the equation refers to the case of particles with equal masses.
The size of the source~$S'$ is of the order of the spatial extent of
reaction.  The~possibility of probing structures of this size arises
when the wave-function modulus squared, $|\Phi^{(-)}|^2$, possesses
pronounced structures, either due to interaction or symmetrization,
that vary rapidly with the relative momentum, typically at low
momenta.
The two-particle cross section can be normalized to the
single-particle cross sections to yield the correlation
function~$C$:
\bed
C(\vec{ p}_1 - \vec{ p}_2) =
{  {d^6 \sigma \over d^3{ p}_{1} \, d^3{ p}_{2}} \over
{d^3 \sigma \over d^3{ p}_{1}} \, {d^3 \sigma \over
d^3{ p}_{2}}} = \int
d^3{ r} \,
S_{\vec{ P}} (\vec{ r}) \,
|\Phi_{\vec{ p}_1 - \vec{ p}_2}^{(-)} (\vec{ r})|^2 \, .
\label{CPS}
\eed
The source~$S$ is normalized to~1 as, for large
relative momenta, $C$ is close to~1 and $|\Phi|^2$ in
(\ref{CPS}) averages to~1:
\bed
\int d^3{ r}
\, S_{\vec{P}} (\vec{ r}) = 1 \, .
\eed

Depending on how the particles are emitted from a~reaction, the
source may have different features.  For a~prompt emission, we expect
the~source to be compact and generally isotropic.  In the
case of prolonged emission, we expect the~source to be
elongated along the pair momentum, as the emitting system moves
in the two-particle cm.  Finally, in the case of secondary
decays, we expect the~source may have an~extended tail.

In the following, I shall discuss restoring the
sources in heavy-ion reactions and extracting information
from the images \cite{bro97}.

\section{Imaging in the Reactions}

The interesting part of the correlation function is its deviation from~1
so we rewrite~(\ref{CPS})
\bea
\nonumber
\Rftn{P}{q}  =
        \Corrftn{P}{q}-1
&=&  \int \dn{3}{r} \left(\wfsquare{{q}}{{r}}-1\right)
        \Source{P}{r} \\
    &=&  \int \dn{3}{r} K(\vec{q},\vec{r})
\, \Source{P}{r}  \, .
\label{RKS}
\eea
From~(\ref{RKS}), it is apparent that to make
the imaging possible $\wfsquare{q}{r}$ must deviate from~1
either on account of symmetrization or interaction within
the pair.  The angle-averaged version of (\ref{RKS}) is
\bed
{\cal R}_{P}({q})  =  4 \pi
\int dr \, r^2 \,
         K_0 ({q},{r}) \,
S^0_P(r)
\label{RKS0}
\eed
where $K_0$ is the angle-averaged kernel.

Let us first take the case of identical bosons with negligible
interaction, such as neutral pions or gammas.  The two-particle
wavefunction is then
\bed
\wftn{q}{r}=\frac{1}{\sqrt{2}}\left( e^{i \vec{q}\cdot\vec{r}}
        +e^{-i \vec{q}\cdot\vec{r}}\right) \, .
\eed
The interference term causes $\wfsquare{q}{r}$ to deviate from 1 and
\bed
K(\vec{q},\vec{r}) = \cos{(2 \vec{q} \cdot \vec{r})}.
\eed
In this case, the source is an inverse Fourier cosine-transform
of ${\cal R}_{P}$.
Also, the angle averaged source can be determined from a~Fourier
transformation (FT) of the angle-averaged~$C$ as the averaged
kernel is
\bed
K^0 (q,r) =
\frac{\sin{(2 q r)}}{2 q r} \, .
\eed

\begin{figure}
\begin{center}
\includegraphics[angle=-90.0,
width=0.68\textwidth]{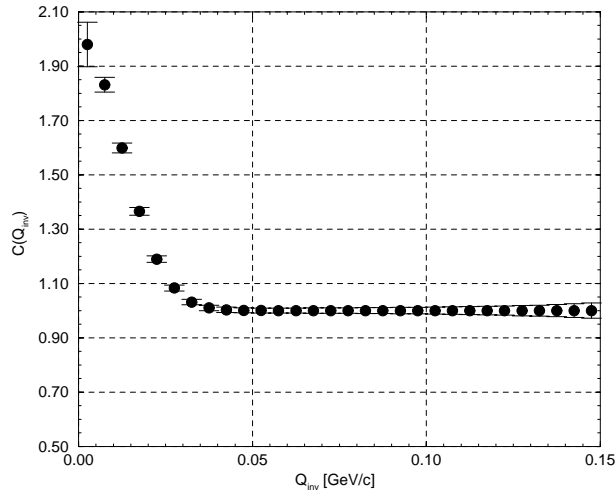}
\end{center}
\caption{Gamow-corrected
$\pi^-\pi^-$ correlation function for Au + Au reaction at
10.8~GeV/nucleon obtained by the E877
collaboration~\protect\cite{bar97}.}
\label{corpi}
\end{figure}

\begin{figure}
\begin{center}
\includegraphics[angle=0.0,
width=0.66\textwidth]{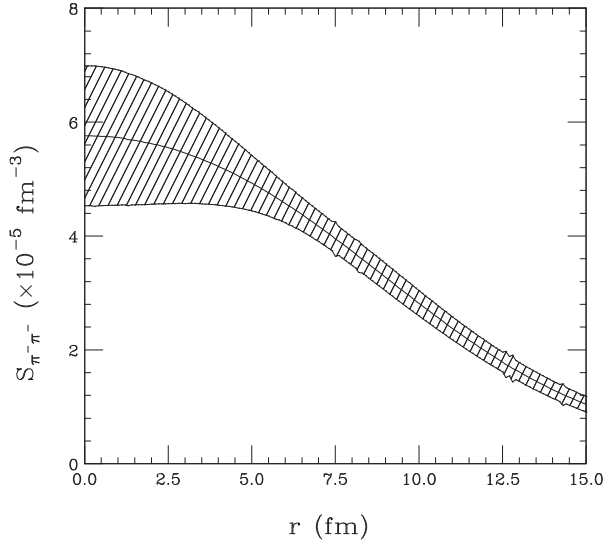}
\end{center}
\caption{Relative source function for negative pions from FT
of the correlation function in Fig.~\protect\ref{corpi}.}
\label{pisor}
\end{figure}

While neutral pion and gamma correlations functions are difficult to measure,
charged pion correlations functions are not.  The charged pion correlations
are often corrected approximately for the pion Coulomb interactions
allowing for the use of FT in the pion source determination.
In Figure~\ref{corpi}, I show one such~corrected correlation function for
negative pions from the Au + Au reaction at 10.8~GeV/nucleon
from the measurements by E877 collaboration at
AGS~\cite{bar97}.
In Figure~\ref{pisor}, I show the relative distribution of emission
points for negative pions obtained through the FT of the
correlation function in Fig.~\ref{corpi}.
The~FT has been cut off at $q_{max} = 200~MeV/c$ giving
a~resolution in the source of $\Delta r \gapproxeq 1/(2 \,
q_{max}) = 2.0$~fm.  The~data spacing gives the highest
distances that can be studied with FT of $r_{max} \lapproxeq 1/(2
\, \Delta q) = 20$~fm.  As you see, the~relative source has
a~roughly Gaussian shape.

\section{Perils of Inversion}

For many particle pairs, such as proton pairs, interactions
cannot be ignored and the straightforward FT cannot be used.
Indeed, even in the charged-pion case, one might want to avoid the
approximate Coulomb correction.  In lieu of this, we can simply
discretize the source and find the source that minimizes the $\chi^2$.
This procedure could work for any particle pair.

With measurements of $C$ at relative momenta $\lbrace q_i
\rbrace$ and assuming the source is constant over intervals
$\lbrace \Delta r_j \rbrace$, we can write Eq.~(\ref{RKS0}) as
\bea
{\cal R}_i =
\Clm{0}{0}{q_i}-1 & = &  \sum_j 4\pi \, \Delta r
\,
              r_j^2 \, K_0 (q_i, r_j) \, S(r_j)\\ &
 \equiv &               \sum_j K_{ij} \, S_j  \, .
\eea
The values ${S_j}$ can be varied to minimize the $\chi^2$:
\bed
\chi^2 = \sum_j  \frac{({\cal R}^{th}(q_j)-
              {\cal R}^{exp}(q_j))^2}{\sigma_j^2 } \, .
\eed
Derivatives of the $\chi^2$ with respect to~$S$ give linear algebraic
eqs. for~$S$:
\bed
 \sum_{ij} {1 \over \sigma_i^2} (K_{ij} \, S_j - {\cal
R}_i^{exp}) \, K_{ij} = 0 \, ,
\eed
with the solution in a schematic matrix form:
\bed
S = (K^\top K)^{-1} \, K^\top \, {\cal R}^{exp} \, .
\label{SKR}
\eed

\begin{figure}
\begin{center}
\includegraphics[angle=90.,width=3.0in]{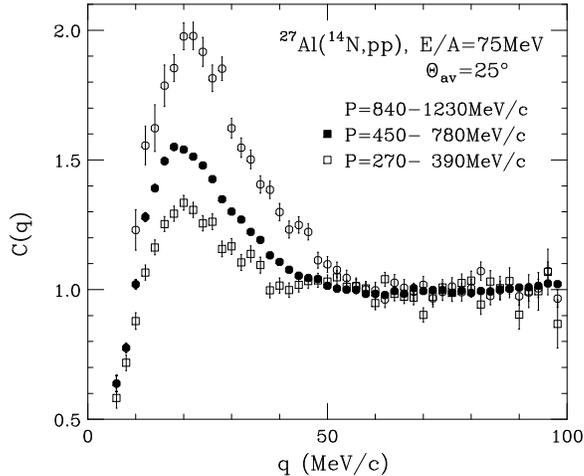}
\end{center}
\caption{
Two-proton correlation function for the $^{14}$N + $^{27}$Al
reaction at~75~MeV/nucleon from the measurements of
Ref.~\protect\cite{gon91} for
three gates of total momentum imposed on protons emitted in the
vicinity of $\theta_{\rm lab} = 25^\circ$. }
\label{gong}
\end{figure}

There is an issue in the above:  how do we discretize the source?
The~FT used before suggests fixed-size bins, e.g.~$\Delta r = 2$~fm.
However fixed size bins may not be ideal for all situations as I
will illustrate using Fig.~\ref{gong}.  This figure shows the $pp$
correlation function from the measurements~\cite{gon91} of the
$^{14}$N + $^{27}$Al reaction at~75~MeV/nucleon, in different
intervals of total pair momentum.
The different regions in relative momentum are associated with
different physics of the correlation function.  For example, the peak
around $q \sim 20$~MeV/c is associated with the $^{1}S_0$
resonance of the wavefunction with a characteristic scale of fm --
this gives access to a~short range structure of the source.
On the other hand, the decline in
the correlation function at low momenta is associated with the
Coulomb repulsion that dominates at large proton separation and
gives access to the source up to (20--30)~fm or more,
depending on how low momenta are available for~$C$.  Should we
continue at the resolution of~$\Delta r \gapproxeq 2$~fm up to
such distances?   No!  At some point there would not be enough many
data points to determine the required number of source values!
Somehow, we should let the resolution vary, depending on the scale
at which we look.

A further issue is that the errors on the source may explode in certain
cases.  The errors are given by the inverse square of the kernel:
\bed
 \Delta^2 S_j = (K^\top \, K)^{-1}_{jj} \, .
\eed
The square of the kernel may be diagonalized:
\bed
 (K^\top \, K)_{ij} \equiv \sum_k {1 \over \sigma_k^2} K_{ki}
\, K_{kj} = \sum_\alpha \lambda_\alpha \, u_i^\alpha \,
u_j^\alpha \, .
\eed
where $\lbrace u^\alpha \rbrace$ are orthonormal and
$\lambda_\alpha \ge 0$; the number of vectors of equals the
number of $r$ pts.  The errors can be expressed as
\bed
 \Delta^2 S_j =  \sum_\alpha {1 \over \lambda_\alpha} \,
u_j^\alpha \, u_j^\alpha  \, .
\eed
You can see from the last equation that the errors blow up,
and inversion problem becomes unstable,
if one or more $\lambda$'s approach zero.  This
must happen when $K$ maps a~region to zero (remember $K =
|\Phi|^2 - 1$), or when $K$ is too smooth and/or too high
resolution is demanded.  A~$\lambda$ close to 0 may be also
hit by accident.

\begin{figure}
\begin{center}
\includegraphics[angle=0.,width=0.72\textwidth]{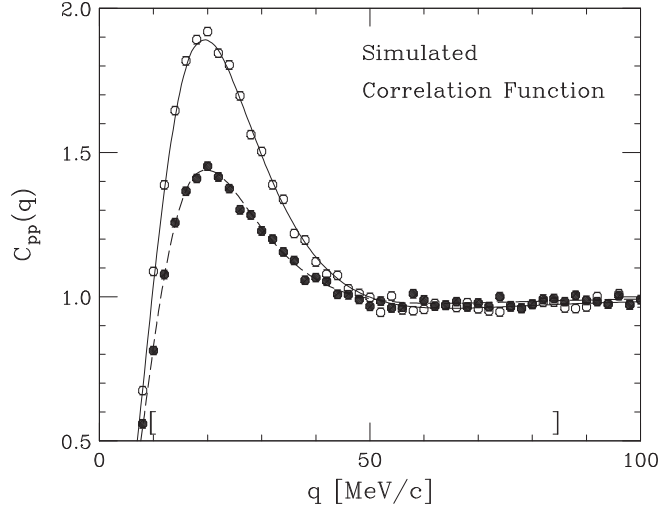}
\end{center}
\caption{
The solid line represents the correlation function from a Gaussian
model source while the dashed lines represent the
correlation functions from a source with an extended tail.
The~points represent values of~$C$ with errors that are
typical for the measurements in Ref.~\protect\cite{gon91}
(Fig.~\protect\ref{gong}). }
\label{simcor}
\end{figure}

\begin{figure}
\begin{center}
\includegraphics[angle=0.,
width=0.72\textwidth]{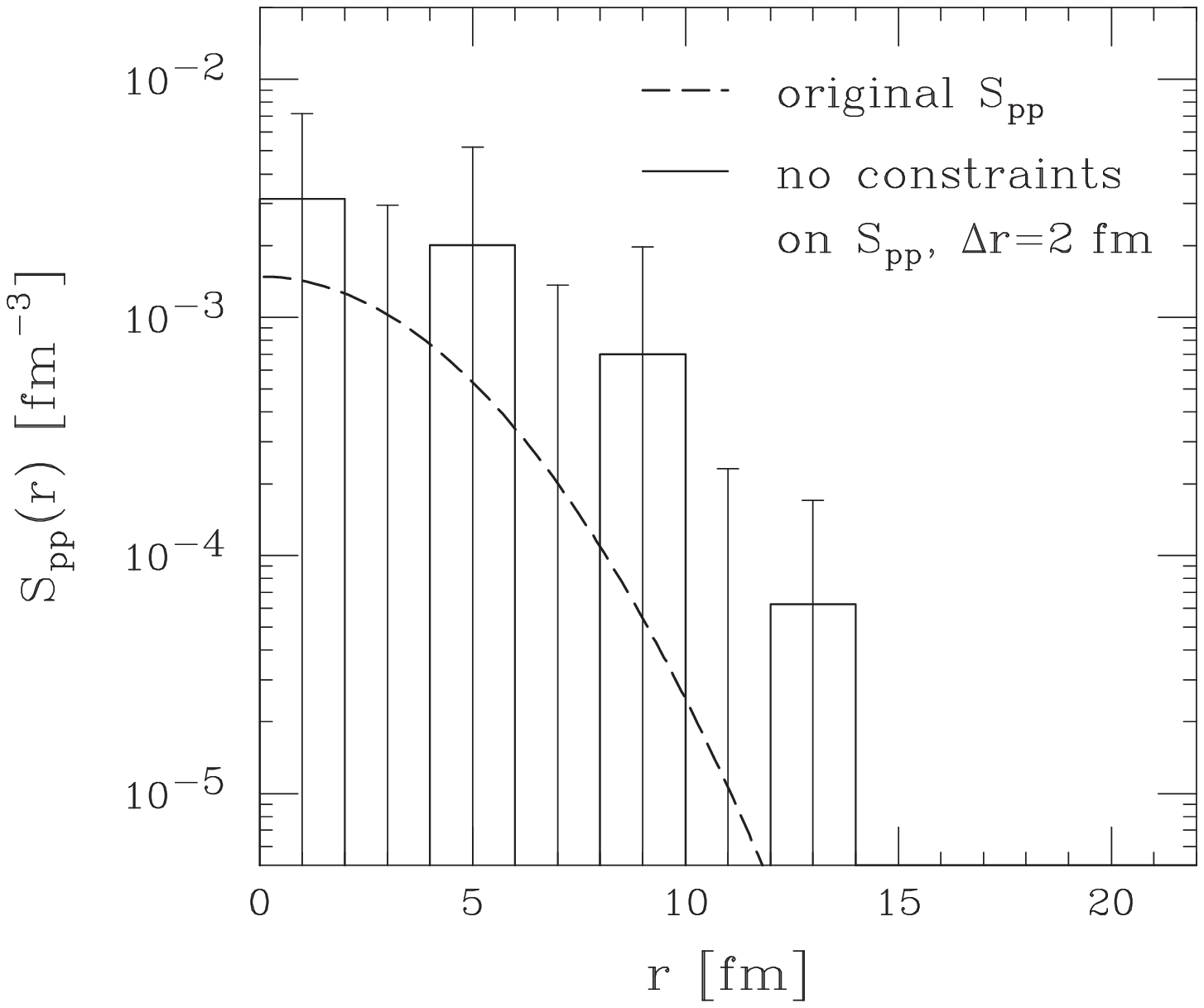}
\end{center}
\caption{
The solid histogram is the relative $pp$ source function $S$
restored from the simulated correlation function in
Fig.~\protect\ref{simcor} from the Gaussian model source
(open symbols there).
The~dashed line is the original source function that we used to
generate the correlation function.  We employed
fixed-size intervals of $\Delta r = 2$~fm and we imposed
no constraints on~$S$.
}
\label{nocon2}
\end{figure}

The~stability issue is illustrated with Figs.~\ref{simcor}
and~\ref{nocon2}.  Figure~\ref{simcor} shows correlation
functions from model sources with small errors added on.
Figure~\ref{nocon2} shows the source in 7~fixed-size intervals
of $\Delta r = 2$~fm.  This source was restored following
Eq.~(\ref{SKR}), from the correlation function indicated in
Fig.~\ref{simcor}.  The~errors in this case far exceed
the original source function.  Every second value of
the restored source is negative.

Vast literature, extending back nearly 75 years, exists on
stability in inversion.  One of the first researchers to recognize
the difficulty, Hadamard, in 1923~\cite{had23}, argued that
the potentially unstable problems should not be tackled.  A~major
step forward was made by Tikhonov~\cite{tik63} who has shown
that placing constraints on the solution can have a~dramatic
stabilizing effect.  In determining the source from data, we
developed a~method of optimized discretization for the source
which yields stable results even without any constraints~\cite{bro97}.

\begin{figure}
\begin{center}
\includegraphics[angle=0.,
width=0.72\textwidth]{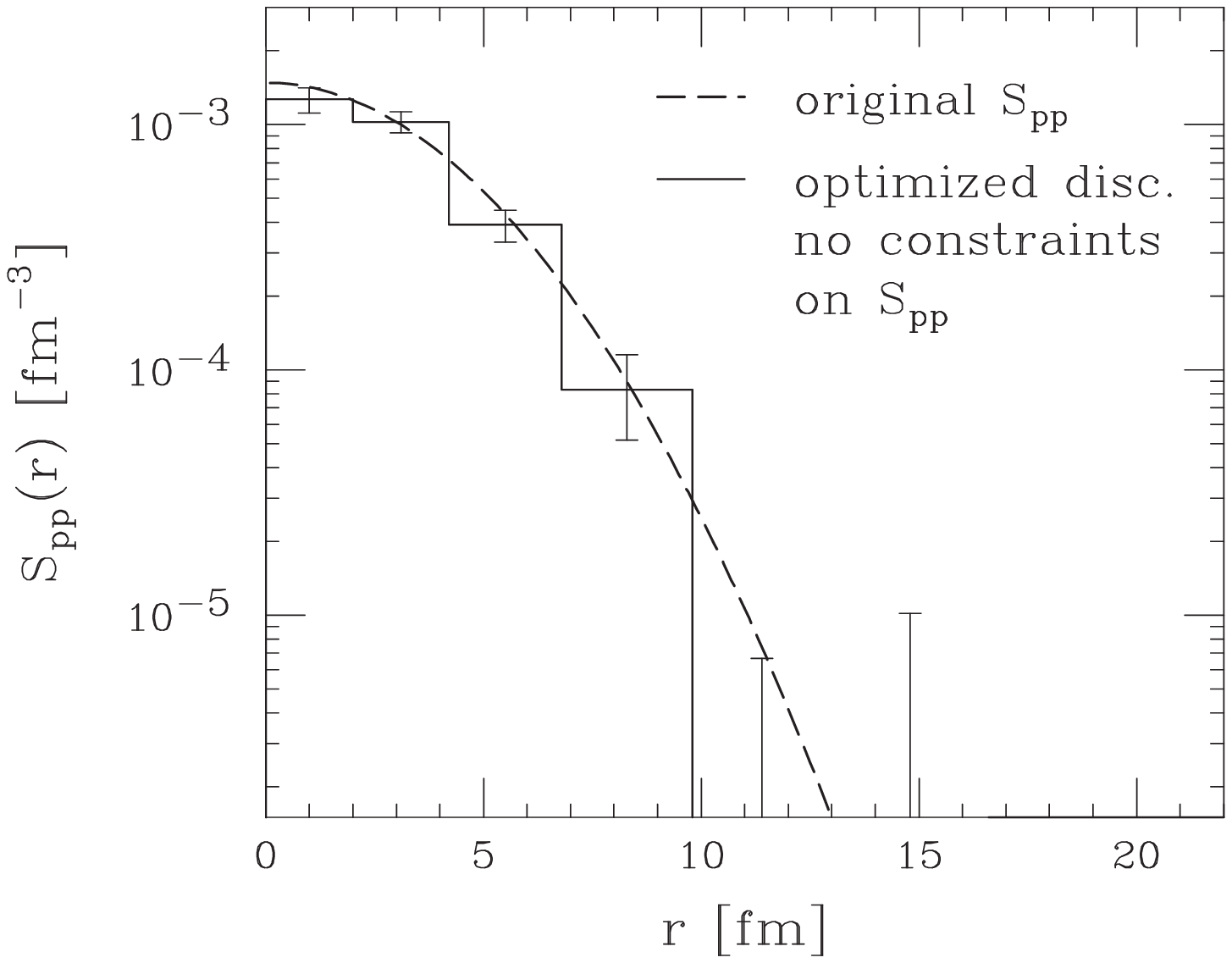}
\end{center}
\caption{
Relative pp source function~$S$ restored (solid histogram)
through the optimized discretization from the correlation
function in Fig.~\protect\ref{simcor} (open symbols there),
together with the original source
function (dashed
line).
}
\label{nocono}
\end{figure}

In our method, we first concentrate on the errors.  We
use the $q$-values for which the correlation function is
determined and the errors of $\lbrace \sigma_i \rbrace$,
but we disregard the values~$\lbrace C_i \rbrace$.  We
optimize the binning for the source function to minimize
expected errors relative to a~rough guess on the source
$S^{mod}$:
\bed
 \sum_j {\Delta S_j \over S_j^{mod}} = \sum_{j} {1 \over
S_j^{mod}} \left( \sum_\alpha
 {1 \over \lambda_\alpha} \,
u_j^\alpha \, u_j^\alpha \right)^{1/2} \, .
\eed
Only afterwards we use $\lbrace C_i \rbrace$ to determine
source values $S_j$ with the optimized binning.  This
consistently yields small errors and an introduction of
constraints may additionally reduce those errors.
The~proton source imaged using the optimized binning from the
correlation function in Fig.~\ref{simcor} is shown
in~Fig.~\ref{nocono}.

\section{pp Sources}

\begin{figure}
\begin{center}
\includegraphics[width=4.55in]{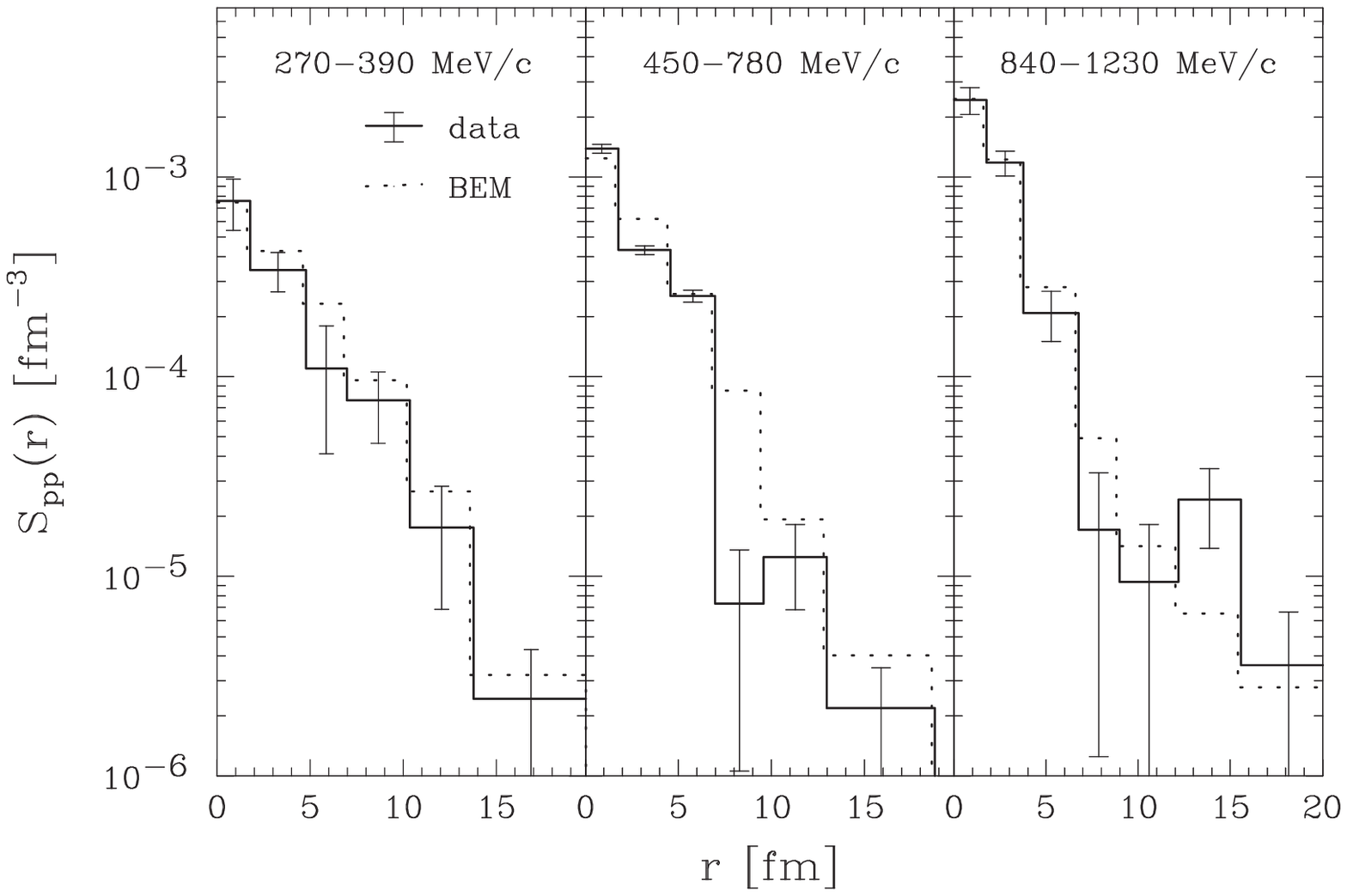}
\end{center}
\caption{
Relative source
for protons emitted from the $^{14}$N + $^{27}$Al
reaction at 75~MeV/nucleon, in the vicinity of $\theta_{\rm
lab} = 25^\circ$, within three intervals of total momentum
of 270--390~MeV/c (left panel), 450--780~MeV/c
(center panel), and 840--1230~MeV/c (right panel).  Solid and
dotted lines
indicate, respectively, the source values extracted from
data~\protect\cite{gon91} and obtained within the
Boltzmann-equation calculation.
}
\label{pps}
\end{figure}

Upon testing the method, we apply it to the~75~MeV/nucleon
$^{14}$N + $^{27}$Al data by Gong \etal~\cite{gon91} shown in
Fig.~\ref{gong}.  In terms of the radial wavefunctions $g$, the
angle-averaged $pp$ kernel is
\bed
 K_0 (q,r)=\frac{1}{2}\sum_{j s \ell \ell'} (2j+1) (g_{j
s}^{j j'} (r))^2-1 \, .
\eed
We calculate the wavefunctions by solving radial Schr\"odinger
equations with REID93~\cite{sto94} and Coulomb potentials.
The~sources restored in the three total momentum intervals are
shown in Fig.~\ref{pps}, together with sources obtained
directly from a~Boltzmann equation model~\cite{dan95} (BEM)
for heavy-ion reactions.

The sources become more focussed around $r=0$ as total momentum
increases.  Now, the~value of the source as $r \rightarrow 0$ gives
information on the average density at freeze-out, on space-averaged
phase-space density, and on the
entropy per nucleon.  The~freeze-out density may be
estimated from
\bed
\rho_{freeze} \simeq N_{\rm part} \times
\Sourcenovec{}{r\rightarrow 0} \, ,
\eed
where $N_{\rm part}$ is participant multiplicity.  Using the
intermediate momentum range, we find
\bed
\rho_{freeze}
\approx
                (17)(.0015{\rm fm}^{-3}) = .16 \rho_0 \, .
\eed
The space--averaged phase-space density may be estimated from
\bed
                f(\vec{p})\approx\frac{(2\pi)^3}{2s+1}\spectra{P}
                \Sourcenovec{\vec{P}}{{r}\rightarrow 0} \, .
\eed
Using the intermediate momentum range we
get $\langle f \rangle \approx .23$ for this reaction.

The transport model reproduces the low-$r$ features of the
sources, including the increased focusing as the total momentum
increases.  The~average freeze-out density obtained directly
within the model is $\rho_{freeze} \simeq .14 \rho_0$.  Despite
the agreement at low~$r$ between the data and the model, we see
important discrepancies at large~$r$.  I discuss these next.

An important quantity characterizing images is the portion
of the source below a~certain distance (e.g.\ the maximum
$r$ imaged):
\bed
\lambda(r_{max})=\int_{r<r_{max}} d^3 r \, S(\vec{r}) \, .
\label{lambda}
\eed
If $r_{max}\Rightarrow \infty$, then $\lambda$ approaches
unity.  The value of $\lambda < 1$ signals that some of the
strength of~$S$ lies outside of the imaged region.  The imaged
region is limited in practice by the available information on
details of~$C$ at very-low~$q$.
We can expect pronounced effects for secondary
decays or for long source lifetimes.  If some particles
stem from decays of long-lived resonances,
they may be emitted far from any other
particles and contribute to $S$ at $r > r_{max}$.

Table~\ref{lpp} gives the integrals of the imaged sources
together with the integrals of the sources from BEM over the
same spatial region.
\begin{table}
\begin{center}
\begin{tabular}{|cr@{$\pm$}lcc|}\hline
\multicolumn{1}{|c}{$P$-Range} &
\multicolumn{3}{c}{$\lambda(r_{max})$} &
\multicolumn{1}{c|}{$r_{max}$} \\ \cline{2-4}
\multicolumn{1}{|c}{[MeV/c]} &
\multicolumn{2}{c}{restored} &
\multicolumn{1}{c}{BEM} &
\multicolumn{1}{c|}{[fm]} \\ \hline
270-390  & 0.69  & 0.15  & 0.98 & 20.0 \\
450-780  & 0.574 & 0.053 & 0.91 & 18.8 \\
840-1230 & 0.87  & 0.14  & 0.88 & 20.8 \\\hline
\end{tabular}
\end{center}
\caption{Integrals of sources from data and BEM in the three
intervals of total momentum.}
\label{lpp}
\end{table}
Significant strength is missing from the imaged sources in the
low and intermediate momentum intervals.  BEM agrees with data
in the highest momentum interval but not in the two
lower-momentum intervals.  In BEM there is no intermediate mass
fragment (IMF) production.  The~IMFs might be produced in
excited states and, by decaying, contribute protons with low
momenta spread out over large spatial distances.  Information
on this possibility can be obtained by examining the IMF
correlation functions.

\section{IMF Sources}

\begin{figure}
\begin{center}
\includegraphics[totalheight=3.3in]{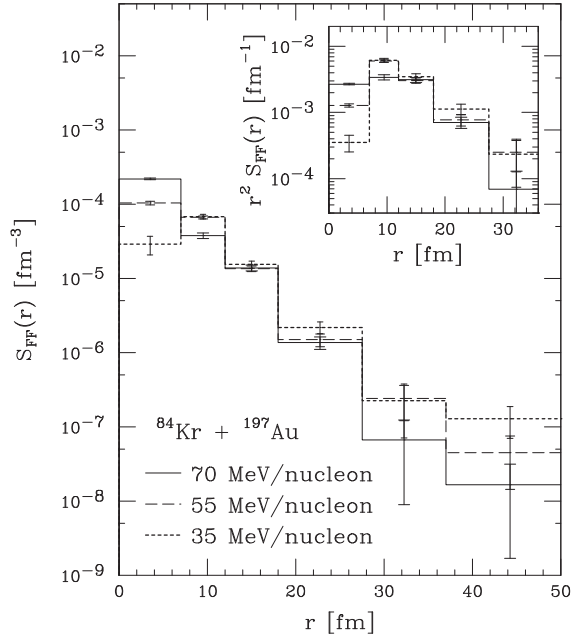}
\end{center}
\caption{
Relative source for IMFs emitted from
central
$^{84}$Kr + $^{197}$Au reactions from the data of
Ref.~\protect\cite{ham96} at 35 (dotted line), 55~(dashed line), and
70~MeV/nucleon (solid line).  The insert shows the source
multiplied by~$r^2$.  In both plots, the full image extends out to $90$~fm.
}
\label{IMF}
\end{figure}

Because of the large charges ($Z \ge 3$), the~kernel in the case
of IMFs is dominated by Coulomb repulsion.  With many
partial waves contributing, the kernel approaches the classical
limit~\cite{kim91}:
\bed
K_0 (q,r)=\theta(r-r_c) (1-r_c/r)^{1/2}-1 \, ,
\eed
where
$r_c=2\mu Z_1 Z_2 e^2/q^2$ is the distance of closest
approach.  There are no IMF correlation data available for the
same reaction used to measure the pp correlation data, so we use
data within the same beam energy range, i.e. the
$^{84}Kr-^{197}Au$ at 35, 55, and 70 MeV/nucleon data
by Hamilton {\em et al.}~\cite{ham96}.
The extracted relative IMF sources are shown in Fig.~\ref{IMF}.
The source integrals for the IMF sources are given in
Table~\ref{IMFt}.  Interestingly, we are nearly capable of restoring
the complete IMF sources.
\begin{table}
\begin{center}
\begin{tabular}{|cr@{$\pm$}lr@{$\pm$}l|}\hline
\multicolumn{1}{|c}{Beam Energy} &
\multicolumn{2}{c}{$\lambda(90 \,{\rm fm})$} &
\multicolumn{2}{c|}{$\lambda(20 \, {\rm fm})$} \\
\multicolumn{1}{|c}{[MeV/A]} &
\multicolumn{2}{c}{ } &
\multicolumn{2}{c|}{ }  \\ \hline
35  & 0.96 & 0.07 & 0.72 & 0.04  \\
55  & 0.97 & 0.06 & 0.78 & 0.03  \\
70  & 0.99 & 0.05 & 0.79 & 0.03  \\\hline
\end{tabular}
\end{center}
\caption{Comparison of the integrals of the midrapidity IMF
source function,
$\lambda(r_{max})$,
in central $^{84}$Kr
+ $^{197}$Au reactions at three beam energies,
for different truncation points, $r_{max}$.
The restored sources use the data of Ref.~\protect\cite{ham96}.}
\label{IMFt}
\end{table}
For the relative distances that are accessible using the pp
correlations ($\sim 20$~fm) we find only (70--80)\% of the IMF
sources.  This is is comparable to what we see for the lowest-momentum
pp source but above the intermediate-momentum proton source.
We should mention that we can not expect complete quantitative
agreement, even if the data were from the
same reaction and pertained to the same particle-velocity
range.  This is due partly to the fact that more protons than final
IMFs can stem from secondary decays.

\section{$\pi^-$ vs. $K^+$ Sources}

We end our discussion of imaging by presenting sources obtained
for pions and kaons from central Au + Au reactions at about
11~GeV/nucleon.  This time we use the optimized discretization
technique rather than
the combination of approximate Coulomb corrections and the FT.
For both meson pairs the kernel $K_0$ is given by
a~sum over partial waves:
\bed
 K_0 (q,r)=\sum_{\ell} \frac{(g^{\ell}(r))^2}{(2\ell+1)} -1 .
\eed
where $g_{\ell}(r)$s stem from solving the radial Klein-Gordon
equation with strong and Coulomb interactions.  In practice the
strong interactions had barely any effect on the kernels and the extracted
sources.

\begin{figure}
\begin{center}
\includegraphics[totalheight=2.70in]{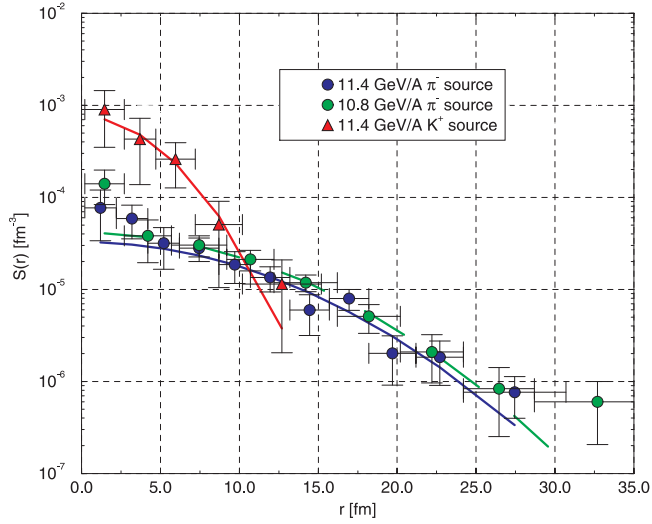}
\end{center}
\caption{
Relative sources of~$\pi^-$ (circles) and of $K^+$ (triangles)
extracted
from central Au + Au data at
11.4~GeV/nucleon~\protect\cite{von98}, for $\pi^-$ and $K^+$,
and at 10.8~GeV/nucleon~\protect\cite{bar97}, for $\pi^-$.
Lines show Gaussian fits to the sources.
}
\label{pikso}
\end{figure}

The data come from the reactions at 10.8~and 11.4~GeV/nucleon~\cite{bar97}.
The respective $\pi^-$ and $K^+$ sources are displayed Fig.~\ref{pikso}.
The kaon source is far more compact than the pion source and
there are several effects that contribute to this difference.
First, kaons have lower scattering cross sections than pions,
making it easier for kaons to leave the system early.  Second,
fewer kaons than pions descend from long-lived resonances.
Next, due to their higher mass, the average kaon has a lower
speed than the average pion, making
the kaons less sensitive to lifetime effects.  Finally, the kaons
are more sensitive to collective motion than pions, enhancing the kaons'
space-momentum correlations.
Differences, qualitatively similar to those seen in Fig.~\ref{pikso},
in the spatial distributions of emission points for kaons and pions
were predicted long ago within RQMD by
Sullivan~\etal~\cite{sul93}.  In the model, they were able to
separate
the different contributions to the source functions.

The~effects of long-lived resonances, mentioned above, are
apparent in the sources extracted from the data.
Thus,
Table~\ref{lampik} gives
\begin{table}
\begin{center}
\begin{tabular}{|cccr@{$\pm$}l|}
\hline
\multicolumn{1}{|c}{} &
\multicolumn{2}{c}{} &
\multicolumn{2}{c|}{} \\[-2.1ex]
\multicolumn{1}{|c}{} &
\multicolumn{1}{c}{$R_0$ [fm]} &
\multicolumn{1}{c}{$\bar{\lambda}$} &
\multicolumn{2}{c|}{$\lambda(35 {\rm fm})$} \\ \hline
$K^+$ (11.4 GeV/A)   & 2.76 & 0.702 & 0.86 & 0.56 \\
$\pi^-$ (11.4 GeV/A) & 6.42 & 0.384 & 0.44 & 0.17 \\
$\pi^-$ (10.8 GeV/A) & 6.43 & 0.486 & 0.59 & 0.22 \\ \hline
\end{tabular}
\end{center}
\caption{
Parameters of Gaussian fits to the sources and integrals
over imaged regions for the central Au + Au reactions.
}
\label{lampik}
\end{table}
the source integrals over the imaged regions together with
parameters of the Gaussian fits to the sources,
\bed
S(r) = \frac{\bar{\lambda}}{(2\sqrt{\pi} R_0)^3} \exp{\left(-\left(
\frac{r}{2 R_0}\right)^2\right)} \, .
\eed
The~errors are quite small for the fitted values.  We find
$\bar{\lambda}_{\pi^-} < \bar{\lambda}_{K^+} < 1$
and~$\bar{\lambda} \lapproxeq \lambda$.

\section{Conclusions}

We have demonstrated that a~model-independent imaging of
reactions is possible.  Specifically, we have carried out
one-dimensional
imaging of pion, kaon, proton, and IMF sources.
The~three-dimensional imaging of pion sources is in
progress.  Our method of optimized discretization allows us to
investigate the sources on a~logarithmic scale up to
large distances.  The sources generally contain information
on freeze-out phase-space density, entropy, spatial density,
lifetime and size of the freeze-out region, as well as
on resonance decays.  The imaging gives us access to the spatial
structure required to extract that information.

\section*{Acknowledgment}
This work was partially supported by the National Science Foundation
under Grant PHY-9605207.


\section*{References}
\begin{thebibliography}{99}


\bibitem{boa90}
D.~H.~Boal, C.~K.~Gelbke, and B.~K.~Jennings,
Rev.~Mod.~Phys.~62, 553 (1990).

\bibitem{bro97}
D.~A.~Brown and \pdd, \plb{398}, 252 (1997);
\prc{57}, 2474 (1998).

\bibitem{bar97}
J.~Barette \etal, \prl{78}, 2916 (1997).

\bibitem{gon91}
W.~G.~Gong \etal, \prc{43}, 1804 (1991).

\bibitem{had23}
J.~Hadamard, {\em Lectures on the Cauchy Problem in Linear
Partial Differential Equations} (Yale U.\ Press, New Haven,
1923).

\bibitem{tik63}
A.~N.~Tikhonov, Sov.\ Math.\ Dokl.\ 4, 1035 (1963).

\bibitem{sto94}
V.~G.~J.~Stoks \etal, \prc{49}, 2950 (1994).

\bibitem{dan95}
P.~Danielewicz, \prc{51}, 716 (1995).

\bibitem{kim91}
Y.~D.~Kim \etal, \prl{67}, 14 (1991).

\bibitem{ham96}
T.~M.~Hamilton \etal, \prc{53}, 2273 (1996).

\bibitem{von98}
T.~Vongpaseuth \etal, Czech.\ J.\ of Phys.\ Suppl.\ S1, 48
(1998).

\bibitem{sul93}
J.~Sullivan \etal, \prl{93}, 3000 (1993).



\end{thebibliography}
\end{document}